\documentclass[aps,prd,floatfix,superscriptaddress,showkeys,longbibliography]{revtex4}
\usepackage{amsmath}
\usepackage{amssymb}
\usepackage{amsfonts}
\usepackage{graphicx}
\usepackage{bbold}
\usepackage{bm}
\usepackage{bm}
\usepackage{cancel}
\usepackage{times,float}
\usepackage{graphicx}
\usepackage[usenames,dvipsnames,svgnames]{xcolor}
\usepackage{slashed}
\usepackage{hyperref}
\hypersetup{colorlinks=true, linkcolor=Blue, citecolor=PineGreen,urlcolor=Blue}
\usepackage{multirow}
\usepackage{ulem}
\usepackage{caption}
\usepackage{subcaption}

\newcommand{\be}{\begin{equation}}
\newcommand{\ee}{\end{equation}}
\newcommand{\bea}{\begin{eqnarray}}
\newcommand{\eea}{\end{eqnarray}}
\newcommand{\beas}{\begin{eqnarray*}}
\newcommand{\eeas}{\end{eqnarray*}}

\begin{document}

\title{Fried-Yennie gauge in pseudo-QED}

\author{Ana Mizher}
\affiliation{Instituto de F\' isica Te\'orica, Universidade Estadual Paulista, Rua Dr. Bento Teobaldo Ferraz, 271 - Bloco II, 01140-070 S\~ao Paulo, SP, Brazil.}
\affiliation{Laboratório de Física Teórica e Computacional, Universidade Cidade de São Paulo, R. Galvão Bueno, 868, Liberdade, 01506-000, São Paulo, Brazil.}
\affiliation{Centro de Ciencias Exactas, Universidad del B\'io-B\'io, Casilla 447, Chill\'an, Chile; ana.mizher@unesp.br}
\author{Alfredo Raya}
\affiliation{Centro de Ciencias Exactas, Universidad del B\'io-B\'io, Casilla 447, Chill\'an, Chile; ana.mizher@unesp.br}
\affiliation{Instituto de F\'{i}sica y Matem\'aticas, Universidad
Michoacana de San Nicol\'as de Hidalgo, Morelia, Michoac\'an
58040, M\'{e}xico; raya@ifm.umich.mx}

\author{Khépani Raya}
\affiliation{ Dpto. Ciencias Integradas, Centro de Estudios Avanzados en Fis., Mat. y Comp., Fac. Ciencias Experimentales, Universidad de Huelva, Huelva 21071, Spain; khepani.raya@dci.uhu.es}

\begin{abstract}

The Fried-Yennie gauge is a covariant gauge for which the mass-shell renormalization procedure can be performed without introducing spurious infrared divergences to the theory. It is usually applied in calculations in regular Quantum-Electrodynamics (QED), but it is particularly interesting to be employed in the framework of pseudo-QED (PQED), where fermions are constrained to 2+1 dimensions while external fields interacting with these fermions live in the bulk of a 3+1 space. In this context, the gauge parameter can be adjusted to match the power of the external momentum in the denominator of the photon propagator, simplifying the infrared region without the need of a photon mass. In this work we apply for the first time this machinery to PQED, generalizing the procedure to calculate the self energy in arbitrary dimensions, allowing of course for different dimensionality of fermions and gauge fields.
\end{abstract}

\keywords{Pseudo-QED, Fried-Yennie gauge, infrared behavior}

\maketitle

\section{Introduction}

For a few decades, the field of condensed matter has provided physical realization of systems that can be associated to manifolds other than the (topologically trivial) 3-space and 1-time dimensions ordinarily considered in high-energy physics. Single-layer materials, quantum wires and nanotubes, are some examples of a plethora of possible arrangements that present alternative dimensionality (see, for instance, ~\cite{Hoffman:ACIE}). Among these, systems arranged in 2-space dimensions have been a hot topic for several years~\cite{Mounet:NatNan}. Surface of liquid helium~\cite{Volovik:ISMP} or the interface of heterostructures are representatives of systems that harbor interesting properties deriving from their space structure. Also, 2D antiferromagnetic insulators~\cite{Lado:PRL121} may give rise to high $T_c$ superconductor~\cite{Dorey:NPB386,Farakos:MPLA13,Franz:PRL87,Herbut:PRB66,Franz:PRB66} and on the other hand (2+1)D electron systems, when placed in a magnetic field, generate new phenomena in the realm of the Quantum Hall effect~\cite{Klitzing:Nat}, such as fractionalization of charge and statistics, statistical transmutation, and so on~\cite{Girving:Book,Ezawa:Book}.

More recently, the discovery of graphene~\cite{Novoselov:science,Zhang:nature,Novoselov:nature}, along with a relatively simple method for its synthesis in the laboratory, have bursted a new interest on (2+1)D systems. After the discovery, it was quickly shown that among several remarkable mechanical and electrical properties, the charge carriers in graphene differ from most condensed matter systems~\cite{Geim:Nature}, exhibiting a quasi-particle behavior that in much resembles relativistic systems, despite having a Fermi velocity around $300$ times lower than the speed of light. This feature is shared with a few of the aforementioned (2+1)D systems but theoretically demonstrated in a clear and simple way for honeycomb lattices. Given this relativistic-like character, it is not surprising that the continuous limit of the tight-binding approach usually applied to describe this systems yields to the Dirac equation in (2+1)D. It was later shown that this framework is suitable to describe a series of other materials (planar or not)~\cite{Mounet:NatNan} and also the later discovered topological insulators~\cite{Hazan:RMP,Qi:RMP,Shankar:preprint}. Due to the common dispersion relation presented by those, they are classified as a family of materials dubbed Dirac materials. 

In general, interaction with fields in general may affect the electronic properties of any material. It is therefore interesting to investigate the effects of interactions of this nature on relativistic-like planar materials. However this kind of study places a challenge since the electrons are constrained to the plane while the dynamical fields are not. This was the main motivation behind the elaboration of pseudo-QED (PQED) \cite{Marino:1992xi}. Specifically, this framework, which is the focus of the present special issue, provides a mixed dimensional theory capable to describe fermionic systems in (2+1)D interacting with dynamical fields in (3+1)D, being them external or generated by the particles themselves. Although it is a non-local theory, it has been shown that it respects unitarity \cite{Marino:2014oba} and causality \cite{doAmaral:1992td}.


In fact, mixed dimensions are intrinsically present in other fields of physics, appearing in approaches like braneworld \cite{Rubakov:2001kp}. Inspired by this framework, a generalization of mixed dimensional theories was developed later being called reduced-QED (RQED). This procedure proposes a treatment to systems of fermions living in generic dimensions different from the gauge fields, as long as the dimension of the former is smaller than the dimension of the latter \cite{Gorbar:2001qt}. RQED has also been widely applied for the particular case of planar materials interacting with dynamical fields. A comprehensive review on this subject, including PQED and RQED main results can be found in \cite{Olivares:2021svj}.

The development of both approaches follows similar paths and consists on dimensionaly reduce the gauge fields, defining an effective theory totally in (2+1)D that accounts for the projection of the gauge field on the plane. This procedure has been widely used to calculate chiral symmetry breaking in planar systems, mainly making use of Schwinger-Dyson techniques \cite{Alves:2013bna,Kotikov:2016yrn,Nascimento:2015ola,Baez:2020dbe,Albino:2022efn}. Renormalization in RQED was investigated in \cite{Teber:2012de,Kotikov:2013eha,Teber:2014hna,Teber:2018goo}, and its scale invariance to all orders was proved in \cite{Dudal:2018pta}. Renormalization group was also applied to investigate the gap in materials like diselenide (WSe$_2$) and molybdenumm disulfide (MoS$_2$) \cite{Fernandez:2020wne}. Aspects of Chern-Simons theory, which is intimately related to PQED were explored in \cite{CarringtonRQED,Olivares:2020eko,Magalhaes:2020nlc,Dudal:2018mms}. Other aspects of the theory like anisotropy in strained graphene \cite{Carrington:2020qfz} and RQED in curved space \cite{RQEDcurved} were analyzed. Effects of parity anomaly associated to a chemical potential were explored in \cite{Dudal:2021ret}.

In the present contribution we will explore the implementation to PQED of a technique that makes the infrared sector of gauge theories more treatable and, although PQED is better behaved in the infrared comparing to ordinary QED, it is still very useful for the regularization of this theory. It consists on performing dimensional regularization in the so-called Fried-Yennie gauge \cite{Fried:1958zz} and manipulating the expressions in a way that the mass-shell renormalization scheme can be implemented without introducing artificial infrared divergences.

This work is organized as follows. In section \ref{sec:FY} we introduce the Fried-Yennie gauge and calculate the self energy in arbritary dimensions, generalizing the formalism (initially applied to ordinary QED$_4$). In section \ref{sec:PQED} we apply the machinery developed in section \ref{sec:FY} to ordinary QED, checking that our approach reproduces the known results, and apply it to PQED. In section \ref{summary} we summarize. Appendices \ref{App:A}, \ref{App:B} \ref{App:C} and \ref{App:D} are dedicated to scrutinize some of the calculations presented in the body of the manuscript.

\section{The Fried-Yennie gauge in D-dimensions}
\label{sec:FY}

\subsection{Setting the stage}

The Fried-Yennie gauge has been explored in the context of quantum chromodynamics~\cite{Boos1988} to explore the quark self-energy with a gauge boson propagator of the form:
\begin{equation}
    \label{eq:PhotonProp}
    D_\beta^{\mu\nu}(k)=-\left(\frac{\lambda_d}{k^2}\right)^{\gamma_d} \left(g^{\mu\nu} + \beta \frac{k^\mu k^\nu}{k^2} \right)\,,
\end{equation}
where $\beta$ is the gauge parameter introduced to ensure transversality in x-space. In the context of mixed-dimensional theories, $\gamma_d$ depends on the space-time dimensionality in which fermions live. The quark propagator is expressed as usual,
\begin{equation}
    S(p)=[\gamma \cdot p -m - \Sigma(p)]^{-1}\,,
\end{equation}
but the self energy is expected to be cast as follows:
\begin{equation}
\label{eq:propExp}
    \Sigma(p)=A+B\,(\gamma\cdot p-m) + C(p)(\gamma \cdot p -m)^2 \;.
\end{equation}
This representation requires $A$ to vanish since the self energy must vanish in the mass-shell $(\gamma \cdot p -m) = 0$, where $m$ is the physical electron mass; also, as discussed in\,\cite{Adkins:1993qm}, $B$ is connected with the electron wavefunction renormalisation constant $Z_2$, via $Z_2=1/(1-B)$. Furthermore, in order for the expansion of Eq.\,\eqref{eq:propExp} to be well defined,
\begin{equation}
    \lim_{\gamma\cdot p \to m} [(\gamma \cdot p - m)C(p)] = 0\;.
\end{equation}
That said, the self-energy in Minkowsky space reads:
\begin{eqnarray}\nonumber
    -i \Sigma(p)&=&i \,\delta m + \int \frac{d^fk}{(2\pi)^d}[-i e(d) \gamma_\mu]\frac{i}{\gamma \cdot (p-k)-m} \\
    \label{eq:SelfEn}
    &\times& [-i e(d)\gamma_\nu] \left[i D_\beta^{\mu\nu}(k)\right]\;.
\end{eqnarray}
Here $\delta_m = m-m_0$ acts as a counterterm ($m_0$ the bare quark mass) and $e(d)=\mu^\epsilon e$ is the electron charge\,\footnote{As usual, $\mu$ and $\epsilon$ are the regulators within a dimensional regularization scheme for $d$-dimensions.}. Here $d$ is the space-time dimension where the fermions live. For notation convenience, in the subsequent, we shall employ Dirac notation $\gamma \cdot p \to \slashed{p}$. Thus, the self-energy reads:
\begin{eqnarray}\label{eq:SelfEn2}
    \Sigma(p)&=&-\delta m + \mathcal{I}\;,\\
    \mathcal{I}&=&
    \int_k \frac{\gamma_\mu [(\slashed{p}-\slashed{k})+m] \gamma_\nu}{(k^2-2 k\cdot p + p^2-m^2)(k^2)^{\gamma_d}} \left(g^{\mu\nu}+\beta \frac{k^\mu k^\nu}{k^2}\right)\,.\nonumber
\end{eqnarray}
where the integration symbol stands for:
\begin{equation}
    \int_k :=(\lambda_d)^{\gamma_d}\left(\frac{\alpha_{em}}{4\pi} \right)  (4\pi \mu^2)^\epsilon \int \frac{d^d k}{i \pi^{d/2}}\;.
\end{equation}
Written in this way, the integral $\mathcal{I}$ in Eq.\,\eqref{eq:SelfEn2} can be split as $\mathcal{I}=\mathcal{I}_F+\beta\mathcal{I}_\beta$, such that:
\begin{eqnarray}
    \mathcal{I}_F&=& \int_k \frac{(2-d)(\slashed{p}-\slashed{k})+ d m}{(k^2-2 k\cdot p + p^2-m^2)(k^2)^{\gamma_d}}\,,\label{eq:DefIF}\\
    \mathcal{I}_\beta&=& \int_k \frac{2 k \cdot p \,\slashed{k}-k^2(\slashed{p}+\slashed{k}+m)}{(k^2-2 k\cdot p + p^2-m^2)(k^2)^{\gamma_d+1}}\,.\label{eq:DefIB}
\end{eqnarray}
Focusing on the former, we introduce a Feynman parametrization to combine the denominators, obtaining:
\begin{eqnarray}\label{eq:IFgen}
    \mathcal{I}_F&=& \int_u \int_k \frac{(2-d)(\slashed{p}-\slashed{k})+ d m}{(k^2- 2u k\cdot p + u (p^2-m^2))^{\gamma_d+1}}\,,\\
    &=& \mathcal{N}_{\gamma_d+1}^{(d)}\int_u [(2-d)(1-u)\slashed{p}+d\,m]\frac{1}{(\mathcal{M}^2)^{\gamma_D}}\;,\nonumber
\end{eqnarray}
where the second line is derived using the formulas from Appendix A. We have adopted the notation:
\begin{eqnarray}\label{eq:intdu}
\int_u &:=&  \gamma_d \int_0^1 du\, (1-u)^{\gamma_d-1}\\
\mathcal{M}^2&:=&u(m^2-(1-u)p^2) \\
\gamma_D &:=& \gamma_d +1 -d/2\, ,\label{eq:gammaD}\\
\mathcal{N}_\alpha^{(d)}&:=&(\lambda)^{\gamma_d}\left(\frac{\alpha_{em}}{4\pi} \right) \frac{\Gamma(\alpha-d/2)}{\Gamma(\alpha)}(-1)^\alpha(4\pi \mu^2)^\epsilon
\end{eqnarray}
A similar procedure is followed for the gauge term contribution, Eq.\,\eqref{eq:DefIB}, thus producing:
\begin{eqnarray}\nonumber
    \mathcal{I}_\beta&=& \frac{\gamma_d+1}{\gamma_d}\int_u \int_k \frac{2(1-u) \, k \cdot p \,\slashed{k}}{(k^2- 2u k\cdot p + u (p^2-m^2))^{\gamma_d+2}}\,,\\
    &-&\int_u \int_k \frac{\slashed{p}+\slashed{k}+m}{(k^2- 2u k\cdot p + u (p^2-m^2))^{\gamma_d+1}}   \label{eq:IBgen1}\;.    
\end{eqnarray}
The second line can be evaluated just as in the $\mathcal{I}_F$ case, while the first, as discussed in Appendix B, requires more attention. We thus have:
\begin{equation}
    \mathcal{I}_\beta=\frac{1}{\gamma_d}\mathcal{N}_{\gamma_d+1}^{(d)}\int_u [\gamma_d m - \{(1-u)(1+\gamma_d-2\gamma_D)\}\slashed{p}]\frac{1}{(\mathcal{M}^2)^{\gamma_D}}
\,.\label{eq:IBgen1}
\end{equation}
Notice that, if $d = d_0 - 2\epsilon$, where $d_0$ is the integer space-time dimension, then (as we shall see)
$\gamma_D = \epsilon$. This result shall be used in the subsequent procedure. Collecting all the terms, one arrives at:
\begin{eqnarray}
    \label{eq:SelfGood}
    \Sigma(p) &=& -  \delta m + \mathcal{N}_{\gamma_d+1}^{(d)}\int_u \Big[\hat{f}_m\, m + \hat{f}_p\,\slashed{p}\Big]\frac{1}{(\mathcal{M}^2)^{\epsilon}}\;,
\end{eqnarray}
where we have defined
\begin{eqnarray}
    \hat{f}_m&:=&d+\beta\,,\nonumber \\
    \hat{f}_p&:=&(1-u)\left((2-d)-\frac{\beta}{\gamma_d}(1+\gamma_d-2\epsilon)\right)
\end{eqnarray}
In the next subsection we discuss how to match the above expression with the representation from Eq.\,\eqref{eq:propExp}.

\subsection{Identifying the dressing functions}
First of all, the counterterm $\delta m$ is fixed by the requirement that $A=0$, arising from the on-shell condition:
\begin{equation}
A= \lim_{\slashed{p} \to m} \Sigma(p)\,,
\end{equation}
which implies
\begin{eqnarray}
\label{eq:counterterm}
    \frac{\delta m}{m} &=& \mathcal{N}_{\gamma_d+1}^{(d)}\int_u \Big[\hat{f}_m + \hat{f}_p\Big]\frac{1}{(u^2m^2)^{\epsilon}}\,\\
    &=&  \bar{\mathcal{N}}_{\gamma_d}\left( 4\pi \frac{\mu^2}{m^2} \right)^\epsilon \frac{\Gamma(1-2\epsilon)\Gamma(\epsilon)}{\Gamma(\gamma_d+2-2\epsilon)} (2\gamma_d + d(1-2\epsilon)) \;, \nonumber
\end{eqnarray}
independent of $\beta$, and where
\begin{equation}
    \bar{\mathcal{N}}_{\gamma_d}=(-1)^{\gamma_d+1} \lambda^{\gamma_d}_d\left(\frac{\alpha_{em}}{4\pi} \right)\,.
\end{equation}

Thus, Eq.\,\eqref{eq:SelfGood} can be rearranged as follows:
\begin{eqnarray}\nonumber
    \Sigma(p) &=& \mathcal{N}_{\gamma_d+1}^{(d)}\int_u \left\{
(\hat{f}_m+\hat{f}_p) m \left[\frac{1}{(\mathcal{M}^2)^\epsilon}- \frac{1}{(u^2m^2)^{\epsilon}} \right] \right. \\
&+&
\left. \hat{f}_p (\slashed p-m) \frac{1}{(\mathcal{M}^2)^\epsilon}\right\} \;.    \label{eq:SelfGood2}
\end{eqnarray}
As can be noted, the second line already displays a $(\slashed{p}-m)$ factor, which will be useful to us when identifying $B$. Furthermore, as discussed in Ref.\,\cite{Adkins:1993qm}, the bracketed term in the first line can be recast so as to extract an overall $(\slashed{p}-m)$ factor as well.

Let $\mathcal{M}^2 = u m^2 \mathcal{G}$, with
\begin{eqnarray}
    \mathcal{G}&:=&u+\kappa(1-u)\,,\\
    \kappa&:=&\frac{m^2-p^2}{m^2}=-\frac{1}{m^2}(\slashed{p}+m)(\slashed{p}-m)\;.
\end{eqnarray}
Then, we can appeal to the following identity:
\begin{equation}
    \frac{1}{\mathcal{G}^\epsilon}-\frac{1}{u^\epsilon} = - \epsilon \kappa(1-u) \int_0^1dv\, \frac{1}{\bar{\mathcal{G}}^{1+\epsilon}}\,,\label{eq:param1}
\end{equation}
with $\bar{\mathcal{G}}=u+\kappa(1-u)v$, such that, the combination of Eq.\,\eqref{eq:param1} and Eq.\,\eqref{eq:SelfGood2} produces:
\begin{eqnarray}
    \label{eq:SelfGood3}
    \Sigma(p) &=& \mathcal{N}_{\gamma_d+1}^{(d)}(\slashed p-m)\int_u  \frac{1}{(um^2)^\epsilon}\left\{\hat{f}_p  \frac{1}{\mathcal{G}^\epsilon}
 \right. \\
&+&
\left. \epsilon(\hat{f}_m+\hat{f}_p) (1-u) \left[\frac{(\slashed{p}+m)}{m}\int_0^1 dv\frac{1}{\bar{\mathcal{G}}^{1+\epsilon}} \right]
 \right\} \;.  \nonumber
\end{eqnarray}
Given the representation from Eq.\,\eqref{eq:propExp}, and the fact that $A=0$, taking the limit $\slashed{p}\to m$ inside the integral yields the value of $B$:
\begin{equation}
    \label{eq:Bexpr}
    B = \mathcal{N}_{\gamma_d+1}^{(d)}\int_u \frac{1}{(um^2)^\epsilon}\left\{\hat{f}_p  \frac{1}{u^\epsilon}+ 2\epsilon(\hat{f}_m+\hat{f}_p)(1-u)  \int_0^1 dv \frac{1}{u^{1+\epsilon}} 
 \right\} \;.
\end{equation}
Performing the evaluation of the integrals, one gets
\begin{eqnarray}
    \label{eq:Bexpr2}
    B &=& -\bar{\mathcal{N}}_{\gamma_d}\left( 4\pi \frac{\mu^2}{m^2} \right)^\epsilon \frac{\Gamma(1-2\epsilon)\Gamma(\epsilon)}{\Gamma(\gamma_d+2-2\epsilon)} \gamma_d(2\gamma_d +d(1-2\epsilon))\nonumber\\
    &=& - \gamma_d \frac{\delta m}{m}\;,
\end{eqnarray}
thus showing a direct relationship with the counterterm $\delta m$. Eqs.\,(\ref{eq:counterterm}, \ref{eq:Bexpr2}) generalize the QED result, presented in \cite{Adkins:1993qm}, for arbitrary dimensions. The substraction of the $B$ contribution to Eq.\,\eqref{eq:SelfGood3} enables us to identify
\begin{eqnarray}
    (\slashed{p}-m)^2C(p) &=& \mathcal{N}_{\gamma_d+1}^{(d)}(\slashed p-m)\int_u  \frac{1}{(um^2)^\epsilon}\\ \nonumber
    &\times&\left\{\hat{f}_p\left[  \frac{1}{\mathcal{G}^\epsilon} - \frac{1}{u^\epsilon} \right] +\epsilon(\hat{f}_m+\hat{f}_p)(1-u)\right. \\
     &\times& \left.\int_0^1 dv \left[\frac{(\slashed{p}+m)}{m}\frac{1}{\bar{\mathcal{G}}^{1+\epsilon}} -\frac{2}{u^{1+\epsilon}}\right] \right\} \nonumber\,.
\end{eqnarray}
Once again, one needs to perform integration tricks in order to extract a global $(\slashed{p}-m)^2$ factor. In the second line, the bracketed term can be expressed in a more useful way by employing the identity from Eq.\,\eqref{eq:param1}. The integrand of the last line, on the other hand, is splitted as follows:
\begin{equation}
     \left[\frac{(\slashed{p}+m)}{m}\frac{1}{\bar{\mathcal{G}}^{1+\epsilon}} -\frac{2}{u^{1+\epsilon}}\right] \to \left[ \frac{\slashed{p}-m}{m}\frac{1}{\bar{\mathcal{G}}^{1+\epsilon}} +2\left(\frac{1}{\bar{\mathcal{G}}^{1+\epsilon}}-\frac{1}{u^{1+\epsilon}} \right)\right]\,,
\end{equation}
so that we have extracted $(\slashed{p}-m)$ for the first term and one can use the following integral identity for the other piece:
\begin{equation}
    \int_0^1 dv \left(\frac{1}{\bar{\mathcal{G}}^{1+\epsilon}}-\frac{1}{u^{1+\epsilon}} \right) = -(1+\epsilon)\kappa(1-u)\int_0^1 dv \frac{1-v}{\bar{\mathcal{G}}^{2+\epsilon}}\;.
\end{equation}
Therefore, one gets:
\begin{eqnarray}
    \label{eq:exprC}
    C(p)&=& \mathcal{N}_{\gamma_d+1}^{(d)} \frac{1}{m} \int_u\int_0^1dv \frac{\epsilon}{(um^2)^\epsilon}(1-u) \\
    &\times&  \left\{ \frac{1}{\bar{\mathcal{G}}^{1+\epsilon}} \left( \hat{f}_p \frac{\slashed{p}+m}{m}+(\hat{f}_p+\hat{f}_m)\right) \right.\nonumber \\
    &+& \left. \frac{2(1-v)}{\bar{\mathcal{G}}^{2+\epsilon}} \left( \hat{f}_p + \hat{f}_m \right)(1+\epsilon)(1-u)\frac{\slashed{p}+m}{m} \right\} \,.\nonumber
\end{eqnarray}
Before discussing the case of $\text{PQED}$, we shall comment on some issues about $C(p)$, related to the convergence of the integral, as well as the selection process of $\beta$ and $\gamma_d$ for a given spacetime dimensionality of the fermion.

\subsection{Scrutinizing $C(p)$: convergence and gauge fixing}
Before going to the particular case $\text{PQED}$, let us further inquire on $C(p)$. For its analysis, it turns out convenient to rewrite Eq.\eqref{eq:exprC} (here $\hat{f}_p := (1-u)\tilde{f}_p$):
\begin{eqnarray}
    \label{eq:exprC2a}
    C(p)&=& \mathcal{N}_{\gamma_d+1}^{(d)} \frac{1}{m} \int_u\int_0^1dv \frac{\epsilon}{(um^2)^\epsilon}(1-u)\Bigg\{\frac{\slashed{p}}{m} \frac{u \tilde{f}_p}{\bar{\mathcal{G}}^{1+\epsilon}} \\
    &+&  \frac{\slashed{p}+m}{m}\left[(1-2u)\frac{1}{\bar{\mathcal{G}}^{1+\epsilon}}-2u(1+\epsilon)(1-u)\frac{1-v}{\bar{\mathcal{G}}^{2+\epsilon}} \right]\tilde{f}_p \nonumber \\
    &+& \left[\frac{1}{\bar{\mathcal{G}}^{1+\epsilon}}+2 \frac{\slashed{p}+m}{m}(1+\epsilon)(1-u)\frac{1-v}{\bar{\mathcal{G}}^{2+\epsilon}}\right](\hat{f}_m+\tilde{f}_p)\Bigg\}\,. \nonumber
\end{eqnarray}
It is argued that the last line of the above expression is not well behaved at small values of $\epsilon$,\,\cite{Adkins:1993qm}. For this reason, and for the simplicity it entails, it is convenient to fix the value of $\beta$ from the requirement that
\begin{equation}
    \hat{f}_m+\tilde{f}_p = \frac{2\gamma_d-\beta(1-2\epsilon)}{\gamma_d}=0\;,
\end{equation}
This constraint leads to a link between $\beta$ and $\gamma_d$,
\begin{equation}
    \label{eq:betaGours}
    \beta(\gamma_d)=\frac{2\gamma_d}{1-2\epsilon}\,.
\end{equation}
which is a particular gauge-fixing for the arbitrary power-like behavior for the photon propagator, in which the transversality condition dictates\,\cite{Boos1988}:
\begin{equation}
    \beta(\gamma_d) = \frac{2\gamma_d}{d-1-2\gamma_d}\;;
    \label{eq:betaGgen}
\end{equation}
in our case, assuming the photon lives in a 4-dimensional spacetime\,\cite{Teber:2012de} (and recalling that $d=d_0-2\epsilon$), 
\begin{equation}
\label{eq:gammaGours}
    \gamma_d = \frac{d_0-2}{2}\;,
\end{equation}
also confirming Eq.\,\eqref{eq:gammaD}, namely $\gamma_D=\epsilon$. For the above, in QED we have $\gamma_d$, such that we recover the well-known result $\beta=2/(1-2\epsilon)$. Also in QED, the second line of Eq.\,\eqref{eq:exprC2a} vanishes, such that $C(p)$ is completely determined by the $\slashed{p}$ term from the first line. This is not the general case where, given Eqs.\,\eqref{eq:betaGours} and \eqref{eq:gammaGours}, one is left with:
\begin{eqnarray}
    \label{eq:exprC2}
    C(p)&=& \mathcal{N}_{\gamma_{d+1}}^{(d)} \frac{\tilde{f}_p}{m^2} \int_u\int_0^1dv \frac{\epsilon(1-u)}{(um^2)^\epsilon}\Bigg\{\slashed{p} u\frac{1 }{\bar{\mathcal{G}}^{1+\epsilon}} \\
    &+&  (\slashed{p}+m)\left[(1-2u)\frac{1}{\bar{\mathcal{G}}^{1+\epsilon}}-2u(1+\epsilon)(1-u)\frac{(1-v)}{\bar{\mathcal{G}}^{2+\epsilon}} \right] \Bigg\}\,. \nonumber
\end{eqnarray}

So, we have properly identified each part of the self-energy: the counterterm (Eq.\,\eqref{eq:counterterm}), the $B$ constant (Eq.\,\eqref{eq:Bexpr2}) and the $C(p)$ dressing function (Eq.\,\eqref{eq:exprC2}). We now turn our attention to the $\text{PQED}$ case.

\section{The Fried-Yennie gauge in PQED}
\label{sec:PQED}

Let us recall that in PQED, implies $d_0=3$ and so: $\gamma_d=1/2$, $\lambda=1/4$ and $\beta=1/(1-2\epsilon)$. Therefore, one has:
\begin{eqnarray}
    \label{eq:CTRQED}
    \frac{\delta m}{m}&=& \bar{\mathcal{N}}_{1/2}\left( 4\pi \frac{\mu^2}{m^2} \right)^\epsilon \frac{\Gamma(1-2\epsilon)\Gamma(\epsilon)}{\Gamma(5/2-2\epsilon)}4(1-\epsilon)^2\;\\
    &\overset{\epsilon\to0}{\approx}& \bar{\mathcal{N}}_{1/2} \frac{16}{3\sqrt{\pi}}\Bigg(\frac{1}{\epsilon}+\Bigg[ \ln \left(4\pi \frac{\mu^2}{m^2}\right) \nonumber \\
    &+&2\frac{\Gamma'(5/2)}{\Gamma(5/2)}-2+\gamma_E\Bigg] \Bigg)+\mathcal{O}(\epsilon)\;,\nonumber
\end{eqnarray}
with $\gamma_E$ the Euler gamma function. According to Eq.\,\eqref{eq:Bexpr2}, $B$ is merely obtained by multiplying the above expression by $-\gamma_d=-1/2$.

Concerning $C(p)$, it is convenient to separate the $\slashed{p}$ and $m$ contributions in Eq.\,\eqref{eq:exprC2} as follows:
\begin{equation}
C(p) = C_p(p) \slashed{p} + C_m(p) m = [C_{p_0}(p) +C_m(p) ]\slashed{p}+C_m(p) m\,,
\end{equation}
where, again, $C_m(p)$ vanishes in QED$_4$. The corresponding integrals are convergent and the limit $\epsilon \to 0$ can be taken safely (see Apendix \ref{App:D}). The final result for the coefficient $C_{p_0}$ in QED$_4$ are: 
\begin{equation}
    C_{p_0}^{QED_4}(p) = \frac{3\bar{\mathcal{N}}_1}{m^2(\kappa-1)^2}\left(-1+\kappa\left(1-\log{\kappa}\right)\right).
    \label{eq:CpQED4}
\end{equation}

Now, focusing on the $\text{PQED}$ case, where $d=3-2\epsilon$ and $\gamma_d=1/2$ we find:
\begin{equation}
 C_{p_0}^{PQED}=   -\frac{16\bar{N}_{1/2}}{3m^2 \kappa\sqrt{\pi}}\left(\frac{5}{3}+{\rm Log}\left[\frac{\kappa}{4}\right]-\ _2F_1^{(0,1,0,0)}\left[2,0,\frac{5}{2},\frac{\kappa-1}{\kappa}\right]\right).
     \label{eq:CpPQED}
\end{equation}
and 
\begin{eqnarray} \nonumber
C_m&=&-\frac{16 \bar{\mathcal{N}}}{3m^2 \kappa \sqrt{\pi}}\Bigg(-2+\left(2-\frac{\kappa}{2}(H_{3/2}+{\rm Log}[\kappa])\right)\\ \nonumber
&-&\frac{3}{2}(2+\kappa)\ _2F_1^{(0,1,0,0)}\left[1,0,\frac{3}{2},\frac{\kappa-1}{\kappa}\right]\ _2F_1^{(0,1,0,0)}\left[1,0,\frac{5}{2},\frac{\kappa-1}{\kappa}\right]\\
&+& (2\kappa) \ _2F_1^{(0,1,0,0)}\left[2,0,\frac{5}{2},\frac{\kappa-1}{\kappa}\right]\Bigg)
\label{eq:CmPQED}
\end{eqnarray}
The full self-energy for PQED in the Fried-Yennie gauge is given by Eq.(\ref{eq:propExp}), where the coefficients can be obtained from  Eq.\,(\ref{eq:CTRQED}),(\ref{eq:CpPQED}) and (\ref{eq:CmPQED}). Note that the coefficient $B$ depends on the mass through ${\rm Log}(\mu^2/m^2)$. After mass renormalization, one can insert this coefficient in the expression for the self-energy and check that the infrared limit yields to a dependence of $m{\rm Log}(\mu ^2/m^2)$ for the second term in Eq.\,(\ref{eq:propExp}). This clearly vanishes in the limit $m\rightarrow 0$. On the other hand, one can easily check that $C_{p_0}^{PQED}$ and $C_m$ go with $1/m^2$. The integrals in Eq.\,(\ref{eq:exprC2}) are not $m$ dependent in the infrared and the only dependence is on the overall $1/m^2$ factor in this equation. When inserted in Eq.(\ref{eq:propExp}), the dependence on the mass in the third term is lifted in the infrared regime and it gives a finite contribution for the self-energy. Under these considerations, it is clear that the self-energy is well-defined in the infrared for any value of the mass, including the limit of vanishing mass.
\section{Summary}
\label{summary}
In this article we have carried out an explicit one-loop calculation of the fermion self-energy in the mixed dimensional theory of Reduced or Pseudo-QED with 4-dimensional photons and 3-dimensional fermions. We have selected to work in the covariant Fried-Yennie gauge and implemented an explicit mass-shell renormalization of $\Sigma(p),$ which acquires the form shown in Eq.~\eqref{eq:propExp}. The present calculation generalizes the previously known case of QED$_4$ carried out in \cite{Adkins:1993qm} motivated by the Coulomb static interaction among charge carriers in low-energy graphene. Although one usually considers graphene as a gapless system, there are a number of proposal for mechanisms which can open the gap and thus induce a mass for electrons in the material, including self interactions~\cite{Gorbar:2001qt,Olivares:2021svj,Baez:2020dbe,Kotikov:2016yrn,Nascimento:2015ola,Alves:2013bna,Albino:2022efn}. In this particular context our approach is suitable for calculations.

The fermion self energy in arbitrary dimensions is defined by the functions in Eq.\eqref{eq:Bexpr2} and~\eqref{eq:exprC2}. For the particular case of PQED these expressions simplify to the ones shown in  Eq.(\ref{eq:CTRQED}), (\ref{eq:CpPQED}) and (\ref{eq:CmPQED}) once dimensionally regularized.  It should be noticed that these expressions do not introduce any spurious infrared divergences whatsoever. Of course, the limit $m\to 0$ is straightforward to obtain and as a result, the self-energy in Eq.~\eqref{eq:propExp} is finite, as $m B(p)$ and $m^2 C(p)$ are finite as $p\to 0$. This is a nice feature of the Fried-Yenni gauge. 

Finally, it is worth mentioning that the gauge here adopted is being considered to calculate the structure the fermion-photon vertex correction to this theory and hence the anomalous magnetic moment of charge carriers in graphene. On the other hand, the massive fermion propagator in arbitrary gauge is also being considered. Results will be reported elsewhere.


\acknowledgments{K. R. wants to acknowledge J. M. Morgado for his valuable help at an early stage of this work. A. R.  acknowledges enlightening discussions with Andrei Davydychev. The work K.~R.~is supported by the Spanish MICINN grant PID2019-107844-GB-C2, and regional Andalusian project P18-FR-5057. AR acknowledges financial support under grant FORDECYT PRONACES/61533/2020.}

\appendix
\section[\appendixname~\thesection]{About the 4-momentum integrals}
\label{App:A}

Recalling that
\begin{equation}
    \int_k := (\lambda)^{\gamma_d}\left(\frac{\alpha_{em}}{4\pi} \right)(4\pi \mu^2)^\epsilon \int \frac{d^d k}{i \pi^{d/2}}\;,
\end{equation}
the following useful formulas have been employed througout this work:
\begin{equation}
    I^S_\alpha(d) := \int_k \frac{1}{(k^2+2 k \cdot v - z)^\alpha} = \mathcal{N}_\alpha^{(d)}\frac{1}{(v^2+z)^{\alpha-d/2}} \label{app:IS}\;,
\end{equation}
\begin{equation}
     I^V_{\mu,\alpha}(d) := \int_k \frac{k_\mu }{(k^2+2 k \cdot v - z)^\alpha}=-v_\mu I^S(d)\;,
      \label{app:IV}
\end{equation}
\begin{eqnarray}
\label{app:IT}
    I_{\mu\nu,\alpha}^T(d)&:=&\int_k \frac{k_\mu k_\nu}{(k^2+2 k \cdot v - z)^\alpha} \nonumber \\
    &=& I^S_\alpha(d)\left(v_\mu v_\nu - \frac{1}{2}g_{\mu\nu}\frac{v^2+z}{\alpha-d/2-1} \right)\;.
\end{eqnarray}
Here, 
\begin{equation}
\label{app:Calpha}
    \mathcal{N}_\alpha^{(d)}:=(\lambda)^{\gamma_d}\left(\frac{\alpha_{em}}{4\pi} \right) \frac{\Gamma(\alpha-d/2)}{\Gamma(\alpha)}(-1)^\alpha(4\pi \mu^2)^\epsilon
\end{equation}

\section[\appendixname~\thesection]{On the $\beta$ integral}
\label{App:B}

Let us consider the gauge term contribution to the self energy, written down in Eq.\,\eqref{eq:DefIB}:
\begin{eqnarray}\nonumber
    \mathcal{I}_\beta&=& \frac{\gamma_d+1}{\gamma_d}\int_u \int_k \frac{2(1-u) \, k \cdot p \,\slashed{k}}{(k^2- 2u k\cdot p + u (p^2-m^2))^{\gamma_d+2}}\,,\\
    &-&\int_u \int_k \frac{\slashed{p}+\slashed{k}+m}{(k^2- 2u k\cdot p + u (p^2-m^2))^{\gamma_d+1}}   \label{app:IBgen1}\;.    
\end{eqnarray}
The evaluation of the second line is straightforward. For the first line, let us consider the momentum integration firstly:
\begin{eqnarray*}
  &&   \int_k \frac{2 k \cdot p \,\slashed{k}}{(k^2- 2u k\cdot p + u (p^2-m^2))^{\gamma_d+2}}\\
    &=&  2 p^\mu \gamma^\nu \int_k \frac{k_\mu k_\nu}{(k^2- 2u k\cdot p + u (p^2-m^2))^{\gamma_d+2}}  \\
    &=& 2 p^\mu \gamma^\nu \left(u^2 p_\mu p_\nu - \frac{1}{2} \frac{\mathcal{M}^2}{\gamma_d+2-1-d/2} g_{\mu\nu}\right) I_{\gamma_d+2}^S(d) \\
    &=&-2\slashed{p}\left(u^2 p^2 - \frac{1}{2}\frac{\mathcal{M}^2}{\gamma_D} \right) \frac{\gamma_D}{\gamma_d+1}\frac{1}{\mathcal{M}^2}I_{\gamma_d+1}^S(d)\\
    &=& 2 \slashed{p}\left(\frac{1}{2}- \frac{u^2 p^2}{\mathcal{M}^2} \gamma_D \right) \frac{1}{\gamma_d+1}\mathcal{N}_{\gamma_d+1}^{(d)}\frac{1}{(\mathcal{M}^2)^{\gamma_D}}\;,
\end{eqnarray*}
where we have employed the formula from Eq.\,\eqref{app:IT}. Therefore, we have:
\begin{eqnarray}\nonumber
    \mathcal{I}_\beta&=& \frac{2}{\gamma_d}\mathcal{N}_{\gamma_d+1}^{(d)}\int_u  (1-u) \left(\frac{1}{2} - \gamma_D \frac{u^2 p^2}{\mathcal{M}^2}  \right)\slashed{p} \frac{1}{(\mathcal{M}^2)^{\gamma_D}}\\
    &-&\mathcal{N}_{\gamma_d+1}^{(d)} \int_u  [(1+u)\slashed{p}+m]\frac{1}{(\mathcal{M}^2)^{\gamma_D}}    
    \label{eq:IBgen2}\,;
\end{eqnarray}
or, more conveniently,
\begin{eqnarray}\nonumber
    \mathcal{I}_\beta&=&\frac{1}{\gamma_d}\mathcal{N}_{\gamma_d+1}^{(d)} \int_u  [\{(1-u)-\gamma_d(1+u)\}\slashed{p}- \gamma_d m]\frac{1}{(\mathcal{M}^2)^{\gamma_D}}\\
    &-& \frac{2}{\gamma_d}\mathcal{N}_{\gamma_d+1}^{(d)}\int_u  (1-u)\gamma_D u^2 p^2 \slashed{p} \frac{1}{(\mathcal{M}^{2})^{\gamma_D+1}}   
    \,.\label{app:IBgen3}
\end{eqnarray}
The second line above can be recast, using integration by parts, in order to have the same power of $\mathcal{M}^2$, thus producing:
\begin{equation}
    \int_u  (1-u)\gamma_D u^2 p^2  \frac{1}{(\mathcal{M}^{2})^{\gamma_D+1}} = \int_u [(1-\gamma_D)-(1+\gamma_d-\gamma_D)u]\frac{1}{(\mathcal{M}^2)^{\gamma_D}}
\,,\label{app:IBgen4}
\end{equation}
which holds for  $\gamma_D \textless 0$ and $\gamma_d \textgreater 0$. Finally, we arrive at a rather compact expression for $\mathcal{I}_\beta$:
\begin{equation}
    \mathcal{I}_\beta=\frac{1}{\gamma_d}\mathcal{N}_{\gamma_d+1}^{(d)}\int_u [\gamma_d m - \{(1-u)(1+\gamma_d-2\gamma_D)\}\slashed{p}]\frac{1}{(\mathcal{M}^2)^{\gamma_D}}
\,.\label{app:IBgen5}
\end{equation}

\section[\appendixname~\thesection]{More about $C(p)$}
\label{App:C}

After performing the integral over $dv$, we can identify the coefficients accompanying $\slashed{p}$ and $m$ contributions to $C(p)$:
\begin{equation}
C(p) = C_p(p) \slashed{p} + C_m(p) m = [C_{p_0}(p) +C_m(p) ]\slashed{p}+C_m(p) m\,,
\end{equation}
where
\begin{equation}
\label{eq:RefCp0}
    C_{p_0}(p) = - \bar{\mathcal{N}}_{\gamma_d} \left( 4\pi \frac{\mu^2}{m^2} \right)^\epsilon\frac{\Gamma(\epsilon)}{\Gamma(\gamma_d)}\frac{\tilde{f}_p}{\kappa}\int_u \frac{u}{(um^2)^\epsilon}\left(\frac{1}{\mathcal{G}^\epsilon}-\frac{1}{u^\epsilon} \right)\,
\end{equation}
and with the $C_m(p)$ dressing function, which vanishes in regular QED, being:
\begin{eqnarray}
\label{eq:RefCm}
    C_{m}(p) &=& \bar{\mathcal{N}}_{\gamma_d}\left( 4\pi \frac{\mu^2}{m^2} \right)^\epsilon \frac{\Gamma(1+\epsilon)}{\Gamma(\gamma_d)}\frac{\tilde{f}_p}{\kappa}\int_u \frac{(1-u)}{(um^2)^\epsilon}\\
    &\times&\Bigg(2u\Bigg[\frac{1}{\mathcal{G}^{1+\epsilon}}-\frac{1}{u^{1+\epsilon}} \Bigg]-\frac{(1-2u)}{(1-u)\epsilon}\Bigg[\frac{1}{\mathcal{G}^\epsilon}-\frac{1}{u^\epsilon} \Bigg]\Bigg)\,.\nonumber
\end{eqnarray}

\section[\appendixname~\thesection]{$C(p)$ integrals}
\label{App:D}

The result of the first integral in Eq.(\ref{eq:exprC2}), that gives the coefficient of $\slashed{p}$ alone, identified as $C_{p_0}$, is given by
\begin{eqnarray} \nonumber
   &&\int_u\int_0^1 dv \frac{(1-u)^{\gamma_d}}{(um^2)^\epsilon} \frac{u}{\bar{\mathcal{G}}^{1+\epsilon}}=\\
  && \frac{(m^2)^{-e}}{\kappa}\Bigg(\frac{\Gamma[2-2\epsilon]\Gamma[\epsilon]}{\Gamma[2+\gamma_d-2\epsilon]}+\frac{\kappa^{-\epsilon}\pi\  {\rm Csc}[\pi \epsilon]\  _2\tilde{F}_1[2-\epsilon,\epsilon,2+\gamma_d-\epsilon,\frac{\kappa-1}{\kappa}]}{\Gamma[\epsilon-1]}\Bigg),
\end{eqnarray}
where $_2\tilde{F}_1[a,b,c,d]$ is the regularized Hypergeometric function and $\Gamma[x]$ is the Gamma function. The limit $\epsilon \rightarrow 0$ can be taken safely, yielding to
\begin{eqnarray} \nonumber
   \mathcal{J}_1&=&{\rm Limit}_{\epsilon \rightarrow 0}\left[\tilde{f}_p\left(\frac{4\pi\mu^2}{m^2}\right)^\epsilon\int_u\int_0^1 dv \frac{(1-u)^{\gamma_d}}{(um^2)^\epsilon} \frac{u}{\bar{\mathcal{G}}^{1+\epsilon}}\right]\\
  \nonumber
 &=&-\frac{(d+2\gamma_d)}{\kappa \Gamma[2+\gamma_d]}\Bigg(-1+H_{1+\gamma_d} +{\rm Log}[\kappa]+\ _2F_1^{(0,0,1,0)}\left[2,0,2+\gamma_d,\frac{\kappa-1}{\kappa}\right]\\ \nonumber
 &-&\ _2F_1^{(0,1,0,0)}\left[2,0,2+\gamma_d,\frac{\kappa-1}{\kappa}\right]+ \ _2F_1^{(1,0,0,0)}\left[2,0,2+\gamma_d,\frac{\kappa-1}{\kappa}\right]\Bigg).
 \end{eqnarray}
Here, $_2F_1^{(i,j,k,l)}[a,b,c,d]$ (without the tilde) gives the derivative of order $i,j,k,l$ of the Hypergeometric function with the respect to variables $a,b,c,d$. We have also used the gauge condition $\beta=2\gamma_d/(1-2\epsilon)$ to obtain the expression above. For real values of the arguments $\kappa$ and $\gamma_d$, the following derivatives of the Hypergeometric function vanish, 
 \begin{eqnarray} \nonumber
 _2F_1^{(0,0,1,0)}\left[2,0,2+\gamma_d,\frac{\kappa-1}{\kappa}\right]=0\\
 _2F_1^{(1,0,0,0)}\left[2,0,2+\gamma_d,\frac{\kappa-1}{\kappa}\right]=0
 \end{eqnarray}
such that the result can be simplified to
\begin{eqnarray} \nonumber
\mathcal{J}_1= -\frac{(d+2\gamma_d)}{\kappa \Gamma[2+\gamma_d]}\Bigg(-1+H_{1+\gamma_d} +{\rm Log}[\kappa]
 - \ _2F_1^{(0,1,0,0)}\left[2,0,2+\gamma_d,\frac{\kappa-1}{\kappa}\right]\Bigg).
 \end{eqnarray}
 In the case of QED, $\gamma_d=1$ and $d=4$ in the limit $\epsilon\rightarrow 0$. Also, the following identity holds:
 \begin{eqnarray}
     _2F_1^{(0,1,0,0)}\left[2,0,3,\frac{\kappa-1}{\kappa}\right]=\frac{1}{2(\kappa-1)^2}\left(1-4\kappa+3\kappa^2+(2-4\kappa){\rm Log}[\frac{\kappa-1}{\kappa}]\right).
 \end{eqnarray}
Considering this, the complete expression for $C_{p_0}^{QED_4}$, including all the coefficients from Eq.\ref{eq:exprC2}, becomes
 \begin{eqnarray} 
C_{p_0}^{QED_4}&=&\frac{\bar{N}_{1}}{m^2}\mathcal{J}_1\\
&=&\frac{3\bar{N}_{1}}{m^2\kappa(\kappa-1)^2}\left(\kappa^2-\kappa -\kappa^2{\rm Log}[\kappa]\right)
\label{eq:cp0qed4}
 \end{eqnarray}
For PQED, $\gamma_d = 1/2$ and $d=3-2\epsilon$, and the final expression, following the same steps used to obtain \ref{eq:cp0qed4} is:
\begin{equation}
 C_{p_0}^{PQED}=   -\frac{16\bar{N}_{1/2}}{3m^2 \kappa\sqrt{\pi}}\left(\frac{5}{3}+{\rm Log}\left[\frac{\kappa}{4}\right]-\ _2F_1^{(0,1,0,0)}\left[2,0,\frac{5}{2},\frac{\kappa-1}{\kappa}\right]\right).
\end{equation}
In the equation above we have benefited from the relation $-1+H_{3/2}=5/3-{\rm Log}[4]$.  The sum of the other 2 integrals in Eq.(\ref{eq:exprC2}), which gives the coefficient of $\slashed{p}+m$, identified as $C_m$ is given by (considering that other derivatives of the Hypergeometric functions that appear in the result vanish for real values of the parameters):

\begin{eqnarray} \nonumber
 &&{\rm Limit}_{\epsilon \rightarrow 0}\left[\int_u\int_0^1 dv (1-2u)\frac{1}{\bar{\mathcal{G}}^{1+\epsilon}}-2u(1+\epsilon)(1-u)\frac{1}{\bar{\mathcal{G}}^{2+\epsilon}}\right]=\\  \nonumber
&& \frac{\gamma_d}{\kappa^2 \Gamma[2+\gamma_d]}\Bigg(-2+\kappa(1-2\gamma_d)+(2+\kappa(\gamma_d-1))(H_{1+\gamma_d}+{\rm Log}[\kappa])\\ \nonumber
&&-(2+\kappa)(1+\gamma_d)\ _2F_1^{(0,1,0,0)}\left[1,0,1+\gamma_d,\frac{\kappa-1}{\kappa}\right]+ (2\gamma_d)\ _2F_1^{(0,1,0,0)}\left[1,0,2+\gamma_d,\frac{\kappa-1}{\kappa}\right]\\
&&+ (2\kappa)\ _2F_1^{(0,1,0,0)}\left[2,0,2+\gamma_d,\frac{\kappa-1}{\kappa}\right]\Bigg)
\end{eqnarray}
For QED$_4$ the coefficient $C_m$ associated to the integral above vanishes. This can be seen, applying the conditions $\gamma_d=1/2$, $d=3-2\epsilon$ and using the relations
\begin{eqnarray}
    _2F_1^{(0,1,0,0)}\left[1,x,2,\frac{\kappa-1}{\kappa}\right]&=&\frac{\kappa-1-{\rm Log}[\kappa]}{\kappa-1}\\
    _2F_1^{(0,1,0,0)}\left[1,x,3,\frac{\kappa-1}{\kappa}\right]&=&\frac{3+\kappa(\kappa-4)+2{\rm Log}[\kappa]}{2(\kappa-1)^2}\\
    _2F_1^{(0,1,0,0)}\left[2,x,3,\frac{\kappa-1}{\kappa}\right]&=&\frac{1+\kappa(3\kappa-4)+2{\rm Log}[\kappa]-4\kappa{\rm Log}[\kappa]}{2(\kappa-1)^2}.
\end{eqnarray}

For PQED the coefficient $C_m$ becomes

\begin{eqnarray} \nonumber
C_m&=&-\frac{16 \bar{\mathcal{N}}}{3m^2 \kappa \sqrt{\pi}}\Bigg(-2+\left(2-\frac{\kappa}{2}\right)\left(H_{3/2}+{\rm Log}[\kappa]\right)\\ \nonumber
&-&\frac{3}{2}(2+\kappa)\ _2F_1^{(0,1,0,0)}\left[1,0,\frac{3}{2},\frac{\kappa-1}{\kappa}\right]\ _2F_1^{(0,1,0,0)}\left[1,0,\frac{5}{2},\frac{\kappa-1}{\kappa}\right]\\
&+& (2\kappa) \ _2F_1^{(0,1,0,0)}\left[2,0,\frac{5}{2},\frac{\kappa-1}{\kappa}\right]\Bigg).
\end{eqnarray}

\end{document}